\documentclass[times, 10pt,twocolumn]{article} 
\usepackage{latex8}
\usepackage{times}

\usepackage{epsfig}
\usepackage{subcaption}
\usepackage{calc}
\usepackage{amssymb}
\usepackage{amstext}
\usepackage{amsmath}
\usepackage{amsthm}
\usepackage{multicol}
\usepackage{pslatex}
\usepackage{cite}
\usepackage{amsfonts}
\usepackage{algorithmic}
\usepackage{graphicx}
\usepackage{textcomp}
\usepackage{flushend}
\usepackage{lastpage}
\usepackage[square,sort&compress,comma,numbers]{natbib}
\usepackage[
bookmarks=false,        				
bookmarksopen=false,					
bookmarksopenlevel=true,
bookmarksnumbered=section,
colorlinks=false,     				
pdfborder={0 0 0}
]{hyperref}
\usepackage{hyperxmp}
\hypersetup{	
	pdflang={en},
	pdfauthortitle={Dr.},
	pdfcontactaddress={Werner-Heisenberg-Weg 39},
	pdfcontactcountry={Germany},
	pdfcontactregion={Bavaria},
	pdfcontactcity={Neubiberg},
	pdfcontactpostcode={85577},
	pdfcontactemail={peter.hillmann@gmail.com},
	pdfdate={2019},
	bookmarksnumbered=true, bookmarksopen=true, bookmarksopenlevel=1,
	breaklinks=true,
	unicode=false,          		
	pdfdisplaydoctitle=true, 		
	pdftitle={NERD: Neural Network for Edict of Risky Data Streams},
	pdfsubject={PoCyMa AI}, pdfauthor={Peter Hillmann},
	pdfkeywords={Peter Hillmann},
	pdfcreator={pdflatex}, pdfproducer={LaTeX with hyperref},
	pdfnewwindow=true,     				
	pdfstartview=Fit,
	pdfcopyright=Copyright \copyright{} 2019 by Peter Hillmann. All rights reserved. 
}

\begin{document}

\title{NERD: Neural \underline{N}etwork for \underline{E}dict of \underline{R}isky \underline{D}ata Streams \vspace*{-0.2cm}}

\author{Sandro Passarelli, Cem G\"{u}ndogan, Lars Stiemert, Matthias Schopp, and Peter Hillmann\\
Faculty of Computer Science,\\ Bundeswehr University Munich,\\ Werner-Heisenberg-Weg 39, 85577 Neubiberg, Germany\\
\{sandro.passarelli; cem.guendogan; peter.hillmann\}@unibw.de;\\ \{lars.stiemert; matthias.schopp\}@localos.io
\vspace*{-0.3cm}
}

\maketitle
\thispagestyle{plain}
\pagestyle{plain}

\begin{abstract}
Cyber incidents can have a wide range of cause from a simple connection loss to an insistent attack.
Once a potential cyber security incidents and system failures have been identified, deciding how to proceed is often complex. 
Especially, if the real cause is not directly in detail determinable. 
Therefore, we developed the concept of a Cyber Incident Handling Support System.
The developed system is enriched with information by multiple sources such as intrusion detection systems and monitoring tools.
It uses over twenty key attributes like sync-package ratio to identify potential security incidents and to classify the data into different priority categories. 
Afterwards, the system uses artificial intelligence to support the further decision-making process and to generate corresponding reports to brief the Board of Directors.
Originating from this information, appropriate and detailed suggestions are made regarding the causes and troubleshooting measures.
Feedback from users regarding the problem solutions are included into future decision-making by using labelled flow data as input for the learning process.
The prototype shows that the decision making can be sustainably improved and the Cyber Incident Handling process becomes much more effective.
\end{abstract}


\Section{Introduction}
Computer-aided security-related attacks on IT infrastructures are steadily rising despite increasing cyber security awareness~\cite{cichonski2012computer,Morgan2018}. 
The average length of stay from the first signs of attacker activity to detection is over 100~days~\cite{Mandiant2018}.
Accordingly the human society is still far from effective IT attack prevention~\cite{Rohmann2017}. 
This is partly due to the fact that the same technology could be used by the attacking and defending actors~\cite{ERPScan2017}.

One example is the increased use of artificial intelligence~(AI) with neural networks in Intrusion Prevention Systems~(IPS) for attack detection of networks and systems~\cite{Dilek2015,7307098}.
Also AI leads to an improved and simplified preparation as well as execution of attacks and supports information gathering. 

Recent trends~\cite{Morgan2018} show an increasing number of attacks, casualties and damages.
However this rising allows the gathering of valuable experience in the fight against cybercrime. 
The resulting knowledge can be used to develop best practice approaches~\cite{AustralienGov2017} through knowledge exchanges between companies and supported dissemination~\cite{IIROC2017}.
This is important to distinguish between security incidents and non-security-related disruptions in IT operations.

\begin{figure}[htb]
	\vspace*{-0.3cm}
	\centering
	\includegraphics[width=0.38\textwidth]{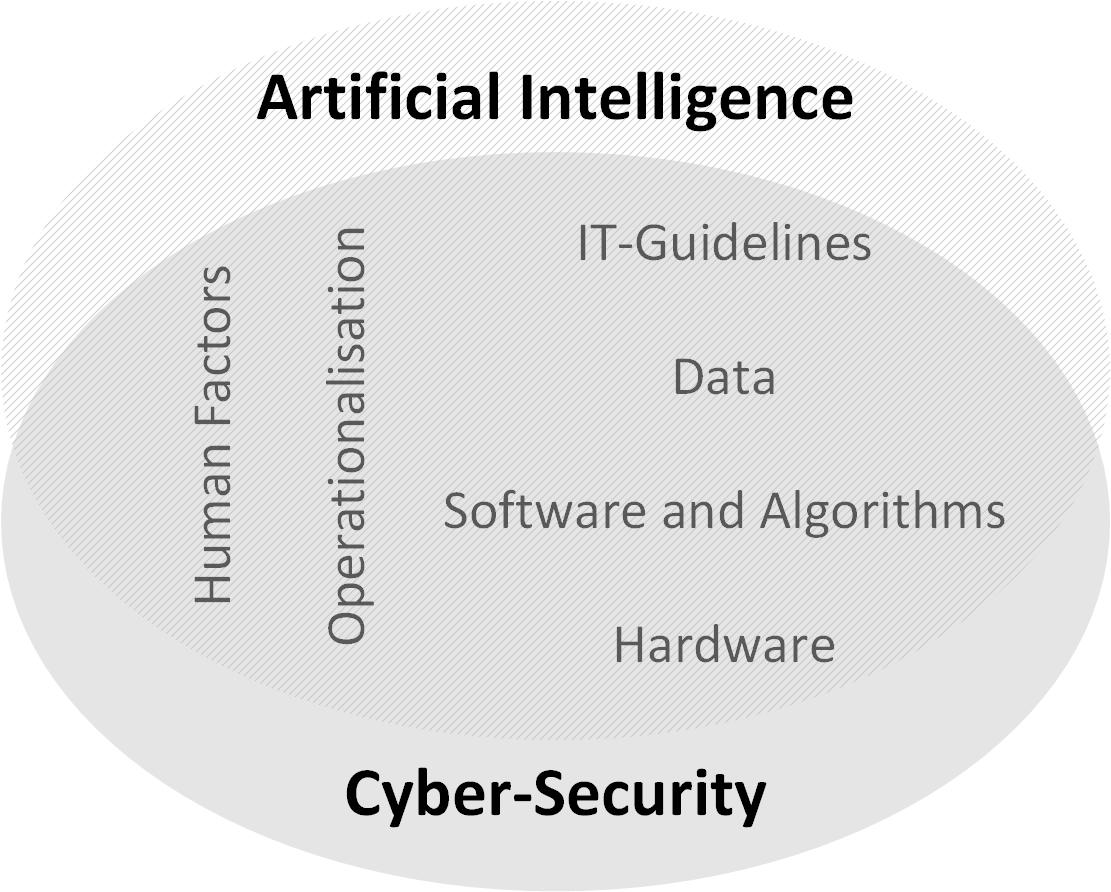}
	\caption[Intersection between AI and Cybersecurity]{Intersection between AI and Cybersecurity~\cite{Zorpette2017}}
	\label{fig:intersection}
	\vspace*{-0.1cm}
\end{figure}

The intersection between AI and Cybersecurity is illustrated by Figure~\ref{fig:intersection}.

Depending on the process, the implementation of the measures extends over several levels from the technician to the management.
Due to different levels of knowledge and responsibilities of decision-makers, communication between organization levels becomes more difficult and can lead to incorrect decisions~\cite{Esseling}.
This happens in particular due to the pressure during an attack.
In such situations, the Computer Security Incident Response Team try to troubleshoot and to contain the attack.
At the same time, the Board of Directors and the management would like to be briefed with non-technical information.
Usually, companies train for such a case for a long time with awareness games and rescue plans to a obtain smooth and felicitous process.
To support the incident handling, this work provides an AI based and management level comprehensive approach against cyber attacks and IT incidents within a company.
This approach goes beyond the standard incident identification and provide the user with concrete and case adapted measures as well as handling activities.

The concept is based on the analysis of available company-internal information from troubleshooting systems and security solutions.
Typical hints comes form intrusion detection systems~(IDS) to propose adequate solutions from several best practices.
Taking into account the resulting benefits, the user can evaluate the solution or add new suggestions or combinations.
This leads in the course of use to a continuous adaptation to the circumstances of the company and its network infrastructure.
With this approach, we attend and realize the suggested process of the NIST Computer Security Incident Handling Guide~\cite{Cichonski2012} with a digital support system.
The focus is especially on the technical phase of Containment, Eradication, and Recovery. 
Furthermore, it helps during the Documentation and Prioritization.
The AI includes the Post-Incident Activity with Lessons Learned.

The remainder of this paper is structured as follows:
Section~2 describes in detail the vision of a AI based Cyber Incident Handling Support System.
This includes a typical scenario combined with a use case and the requirements for such a support system.
Afterwards, Section~3 provides an overview of other approaches in that application area.
The main part in Section~4 presents in detail the scientific approach and method.
We explain the concept from an Enterprise System Engineer's point of view to support decision making during IT incidents with machine learning.
It is integrated in the larger Cyber Incident Response System~\cite{PoCyMa2018}.
Subsequently, we show the evaluation of the running system in Section~5.
Finally, the last section summarizes our work and provides and outlook.

\Section{Scenario and Requirements}
The following scenario illustrates the purpose of the application.
In addition, some scientific questions are called to be answered.
This results in different requirements for a decision support system for a Cyber Incident Handling Support System.

\SubSection{Scenario}
A medium-sized company which uses trouble\-shooting, monitoring, and security systems is regarded as an application user.
Failures and malfunctions of the IT systems lead to direct implications for the company's value creation process.
Therefore, a fast recovery to a normal status of operation is highly important.
Due to the legal situation e.g. the obligation to report security-critical incidents~\cite{Bundestag2015}, the management is confronted with the topic of IT security.

In context of Enterprise Architecture, the management wants to be informed about detected incidents and possible risks as well to be involved in the solution process.
In view of the often narrow financial resources for IT security, existing systems and information such as IDS should be used and integrated into the overall concept of the new AI-based security solution.
The resulting data must be prepared and analysed in the overall context so that they eventually serve as the basis for best-practice proposed solutions and can also be explained in an understandable way to people with less technical affinity.
Since best practices are only indicative and not applicable to all situations, a way to incorporate feedback must be provided.

The following questions arise from these preconditions with regard to the conceptual design of data classification, pattern recognition, decision support, and the usage of feedback: 
\begin{itemize}
	\item[1)] How can an AI be used to enable and continuously improve decision support in the area of incidents?\\ \textit{This question deals with the question which AI approaches are suitable to fulfill the required characteristics by means of the data generated during incidents of analysis tools.}
	\item[2)] How must data be classified and prepared in order to flow recursively into the AI?\\ \textit{The main focus here is on how the generated data and information can be processed for the implementation of the training in order to enable a successful training of the selected AI approach for the selected task area and the identified situations.}  
	\item[3)] How can the AI through recursive learning be personalised in order to adapt to specific circumstances?\\ \textit{This question relates to the premise of the concept of changing proposed solutions by using the user feedback in order to produce a personalized result that includes the correspondingly used network environment. }
\end{itemize}

\SubSection{Requierements}
The previous application example and the scientific questions results in different requirements on decision support systems for IT incidents:
\begin{itemize}
	\item \textit{Analysis of data:} Processing large and different data types leads to the challenges of Big Data.
	\item \textit{Pattern Recognition:} In connection with the data analysis adequate solutions are to be provided by specifying of recognized incidents based on known values and correlations. 
	\item \textit{Decision and command support:} The core functions of the system are to show solutions and multiple possibilities to act during detected incidents. These suggested solutions are intended to provide an overview of detailed and adequate measures that the user can take to counteract an attack and to recover.
	\item \textit{Feedback embedding:} The ability to evaluate individual solution proposals by user feedback to treat future events in a better way.
	\item \textit{Anonymisation:} With the new EU law of the \textit{General Data Protection Regulation}~\cite{Parliament2018}, personal data is suitable to be anonymized in order to prevent possible conclusions about company-specific architectures or processes.
	\item \textit{Extensibility:} The integration of new functions should be possible at any time. Errors and maintenance should be easily identified and improved during operation in order to save resources.
	\item \textit{Automation:} The user should be able to generate a report for the participation of the higher company levels. This should show the management level recognized situations and describe their effects on the IT infrastructure.
\end{itemize}
Further requirements regards the anonymization of personal data, expandability of the system and automation for report generation.

\Section{Related Work}
In order to obtain an overview of the current state of affairs, the systematic method of literature research is used.
According to the classification of Buczak~\cite{7307098}, which is based on the taxonomy of machine learning methods, previous work can be classified as Decision Support~(DS) or Intrusion Detection~(ID).

Currently IBM developes an ID system called QRadar Security Intelligence~\cite{IBM2019}. 
This SIEM uses the AI to improve the performance of business process.
Among other things QRadar supports only IBM Watson.
However this proprietary system does not generate any appropriate suggestions to solve the problem.

Another system in the field of ID is SAP Enterprise Threat Detection~\cite{SAP2019}.
The SIEM focuses only on the detection of threats in the area of Enterprise Architecture, without using a learning system.

The work~\cite{Reyes2019} introduces a system for automated report generation for university CERT teams. However this work does not provide proposals for measures for problem solving. It also does not provide attack detection. The system is just processing data and this is why the concept is to be placed in category DS. In addition, there is no feedback control to improve the system.

The prototype~\cite{7561088} classifies network traffic and serves the DS.
Also the prototype shows the usability of artificial neural networks to improve an IDS. As well as the system FASTT~\cite{Kalyani2017}, it categorizes IDS alarms and groups them together. FASTT uses an AI to improve the DS. Nevertheless, both systems do not provide any recommendations for action.

The study~\cite{Rauchecker2014} introduces a DS-System for IT Security and Incident Management. It aims on the report of detected incidents to the IT security personnel of the first-level support by using operation research. It is similar to the system Scyllarus~\cite{Heimerdinger2003}. Scyllarus combines a bunch of IDS alarms by using a Bayesian network. A disadvantage is that the integration of measures and the feedback loops are missing.

All in all, no approach complies with the identified requirements.

\Section{Concept of the Incident Support Handling System}
The following concept is part of the \textit{PoCyMa} project~\cite{PoCyMa2018}. 
It is a Cyber Incident Response System, which includes multiple sources like vulnerability databases to be aware of up-to-date attack vectors and geo-databases for IP-Geolocation to identify attack sources.
Nevertheless, this presented approach is also usable as stand alone.
The working title PoCyMa \textit{(PotatoCyberMap)} originate from a live attack event on our twitter stream from a person alias cyber potato during our security conference in 2016.
After this event, we started to develop a comprehensive countermeasurement, which this approach is part of it.

\begin{figure*}[htb]
	\centering
	\includegraphics[width=1.0\textwidth]{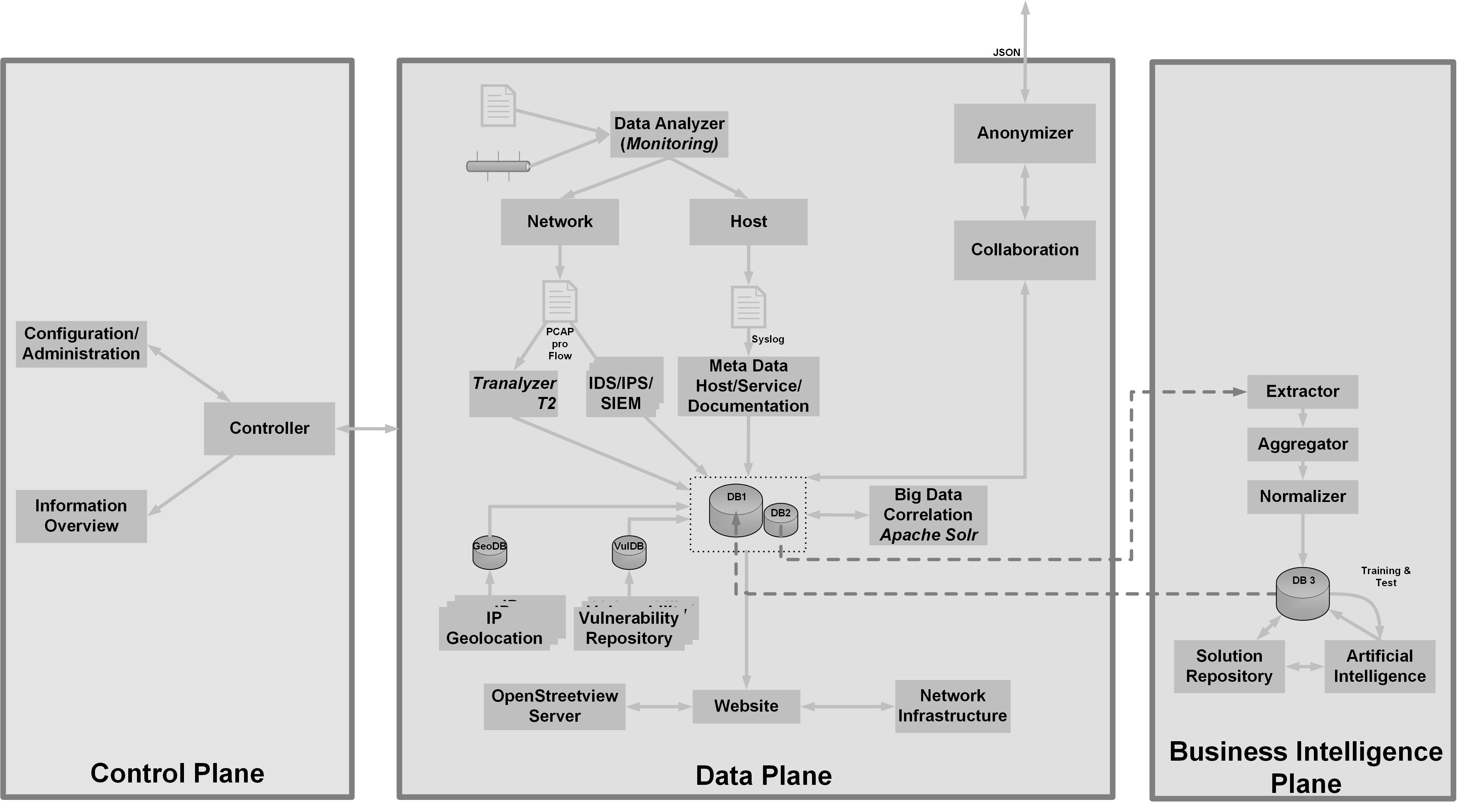}
	\caption{Entire Enterprise Architecture of the Cyber Incident Response System PoCyMa}
	\label{fig:PoCyMa-Entire}
	\vspace*{-0.2cm}
\end{figure*}

\SubSection{Placement inside of the PoCyMa Architecture}
The entire PoCyMa architecture is presnted in \mbox{Figure~\ref{fig:PoCyMa-Entire}.}
Based on the Model-View-Controller principle, the PoCyMa architecture consists of the three planes, being the Data-, Control, and Business Intelligence~(BI) Plane.
Every plane is responsible for a defined task.
The Control Plane is used to monitor and configure the usage of the program.
The Data Plane is used to prepare and bundle data from different sources like data of IDS, SIEM, geolocation, and vulnerabilities with tools like Nagios and Tranalyzer~\cite{7849909} for flow information.
These are used for visualization in respect to the different hierarchy levels.

The NERD project resides inside the BI plane, which is used to support AI-based solutions.
It uses information from troubleshooting systems, self developed rescue plans, and monitoring information.
Furthermore, it uses the internally provided information of the data plane.
These big data are analyzed and correlated by the NERD project to suggest activities according to identified incidents.
Due to logging of events with a ticket system and wiki as well as their respective solutions, there is a broad knowledge base to use.
The overlying objective is to provide suggestions for adequate solutions, based on best practices and solved events.

\begin{figure}[hbt]
	\centering
	\includegraphics[width=0.49\textwidth]{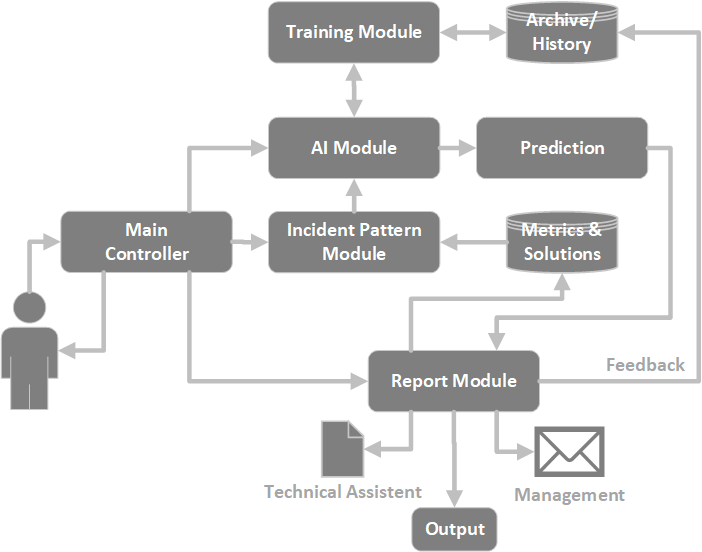}
	\caption{Modularized structure of NERD}
	\label{fig:gesamtkonzept}
	\vspace*{-0.4cm}
\end{figure}

\begin{figure}[hbt]
	\centering
	\includegraphics[width=0.49\textwidth]{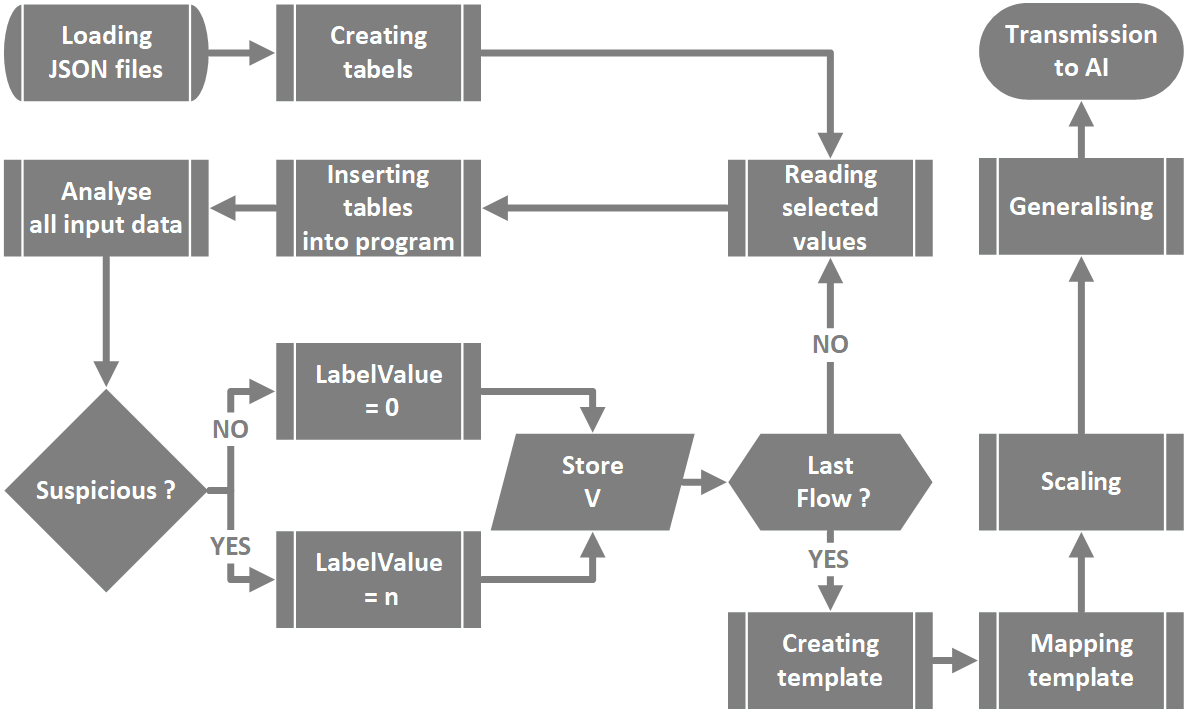}
	\caption{Procedure for preparing the data as SDL diagram}
	\label{fig:Aufbereitung}
	\vspace*{-0.2cm}
\end{figure}

\SubSection{Structuring the System}
The system is built with a modular approach to ensure both, an easy maintainability and expandability.
It is divided into nine modules, which are shown in Figure~\ref{fig:gesamtkonzept}, the procedure of the AI data treatment is shown in Figure~\ref{fig:Aufbereitung}.


The main module provides the main interface for the user and serves as the controller of PoCyMa BI plane. The AI consists of a fully connected \textit{Deep Feed Forward Neural Network} (NN) which uses \textit{two hidden layers} and gets trained via \textit{supervised learning}. The NN is represented by the AI-Module in Figure~\ref{fig:gesamtkonzept}. The \textit{four layers} of the NN are needed to ensure the correct handling of the variable complexity. A \textit{leaky rectified linear unit algorithm}, shown in Formula~\ref{eq:RELU}, is used as the activation function inside the NN. The functionality is based on the classical maximum principle to identify thresholds of the key attributes of the input values, symbolized with \textit{x}.

\begin{figure}[hbt]
	\vspace*{-0.1cm}
	\begin{minipage}[h]{1.0\linewidth}
		\centering
		\begin{equation}\label{eq:RELU}
		f(x) = max(x,0)
		\end{equation}
	\end{minipage}
	\vspace*{-0.3cm}
\end{figure}

The output layer of the NN uses a standard \textit{softmax function}, shown in Formula~\ref{eq:Softmax}, which helps dividing the probability onto the different possible outcomes \textit{z}.
It is typically used for scenarios with multiple classification and regression models as we focus on. The variable \textit{z} represents the input value from a specific neuron j. Each input value of the multiple neuron (1 to K) influences the output value of the considered and firing neuron.
\begin{figure}[hbt]
	\vspace*{-0.6cm}
	\begin{minipage}[h]{1.0\linewidth}
		\centering
		\begin{equation}\label{eq:Softmax}
		\sigma (z)_{j}=\frac{e^{z_{j}}}{\sum_{k=1}^{K}e^{z_{k}}} \text{ for  j=1,...,K}
		\end{equation}
	\end{minipage}
\end{figure}

Training optimization is realised by using the \textit{Adaptive Moment Estimation~(ADAM)}. The train module includes the data analysis and data processing in accordance with the \textit{Extract-Transfer-Load~(ETL)} process~\cite{Strohmeier2008} and therefore handles the preparation of the input data, which are used by the AI and the train module.

The prediction module analyses new data in coordination with the AI module and puts out weighted treatment options, out of an array of trained solutions. Those solutions can be printed via the output module and exported as PDF file.

To ensure a continual improvement of the suggested solutions and treatment options, the given user feedback is used inside the training process. This feedback contains a valuation of the given solutions in respect to their usability and how well the solution fit the initial problem or event. With these user provided inputs, new training sets get created which are then used to further train the NN, which leads to helpful solutions being suggested more often and to a more and more personalized NN.

\SubSection{Data Analysis and Treatment}
According to the structure of PoCyMa, the needed data is getting extracted from the data plane. This data is flow-based and contains information from at least one IDS.
According on this data, a first analysis in regards of certain incidents and attack vectors, as well as a first flow labelling is being made. Non-relevant flows are filtered out of the data set.

The flow data are then transformed into values between \textit{zero} and \textit{one} by using an arithmetic function, so that they can be used in the training process.
The value \textit{zero} represents a normal behaviour and the value \textit{one} shows a abnormal or suspicious behaviour. This arithmetic function varies in respect to the transformed data type and its minimum and maximum value.

After the data has been properly labelled and transformed, it is transferred to the AI module according to the load step of the ETL process. At this time, the data set contains 23 different attributes, which are used by the NN.
Table~\ref{tab:attirbutes} show the different key attributes as input data. These are not limited and can be extended with geoinformation and vulnerability codes.

\begin{table}[hbt]
	\centering
	\caption{Overview of the key attributes for the analysis and classification to suggest handling activities.}
	\label{tab:attirbutes}
	\begin{tabular}{r|l|l}
		ID  & Attribute        & Description                                            \\ \hline
		1     & Flowindex        & ID of concatenated flows                      \\
		2     & Duration         & Timelength of a flow                                   \\
		3     & IP Destination   & IP Address of destination                          \\
		4     & Source Port      & Source Port of sender                              \\
		5     & Destination Port & Destination Port\\&&  of the receiver                       \\
		6     & L4 Protocol      & Protocol of transport layer                        \\
		7     & DstPortClass     & Classification of the port                             \\
		8     & TCP-Rate         & Relation between\\&&  TCP-Packets and\\&&  all Packets of a flow \\
		9     & TCPPAckCntAsm    & TCP-ACK-Asymmetrie                                     \\
		10    & PktAsm           & Paket-Asymmetrie                                       \\
		11    & BytAsm           & Byte-Asymmetrie                                        \\
		12    & TCPStat          & TCP Status                                             \\
		13    & IPMinTTL         & Minimum IP Time to live                                \\
		14    & IPMaxTTL         & Maximum IP Time to live                                \\
		15    & PerPS            & Relation between packets\\&&  send per Flow                 \\
		16    & TCPSeqFCnt-Rate  & Relation between\\&&  TCPSeqFaultCnt\\&&  and TCPSeqCnt          \\
		17    & TCPAckFCnt-Rate  & Relation between\\&&   TCPAckFaultCnt\\&&  and TCPAckCnt         \\
		18    & EstBwPFlow       & Average Bandwith of a flow                             \\
		19    & TCPAggrFlags     & TCP-Flags of a flow                                    \\
		20    & TCPAggrAnomaly   & Aggregated TCP-Header,\\&&  Anomaly-Flags                    \\
		21    & TCPAggrOptions*  & Aggregated  TCP Options                                \\
		22    & TCPStates        & States of a TCP connection                             \\
		23    & Label            & Scenario based label\\&&  including feedback               
	\end{tabular}
\end{table}
For each flow of data, the values are calculated according to the attributes of the table.
As a measure, scaling and generalization are necessary. Metrics defined in a metrics collection are used for this purpose. Since the AI is a neural network, floating point numbers between \textit{zero} and \textit{one} are required.
The calculation of every key attribute is similar for all input values. A special form have the discrete values like the port number.
The index value of the current attribute is divided by the maximum index number defined by the standard or request for comments~(RFC).
Subsequently, a two-dimensional mapping of the attributed flows to the input data template of the processing pipeline takes place.
Nevertheless, each flow is analysed individually, before a correlation between multiple flows takes place.


\SubSection{Decision Support and Feedback Usage}
The NN is constructed in a way, that it can create treatment recommendations based on the enriched flow data, by classifying the data after they were analysed and prepared by the ETL process. 
For the first prototype, we limited the classification to normal traffic, service incident, and denial-of-service attack.
According to these classifications, the user gets shown an array of treatment recommendations, which are based on known best practices and information from solved events.
These solutions can then be rated by the user in respect to their efficiency.
This rating gets included into the training of the NN by using an updated training set, which considers the before made ratings of the user, for every training process that occurs.
This contributes to the aforementioned continual solution improvement. 

\begin{figure}[hbt]
	\centering
	\includegraphics[width=0.49\textwidth]{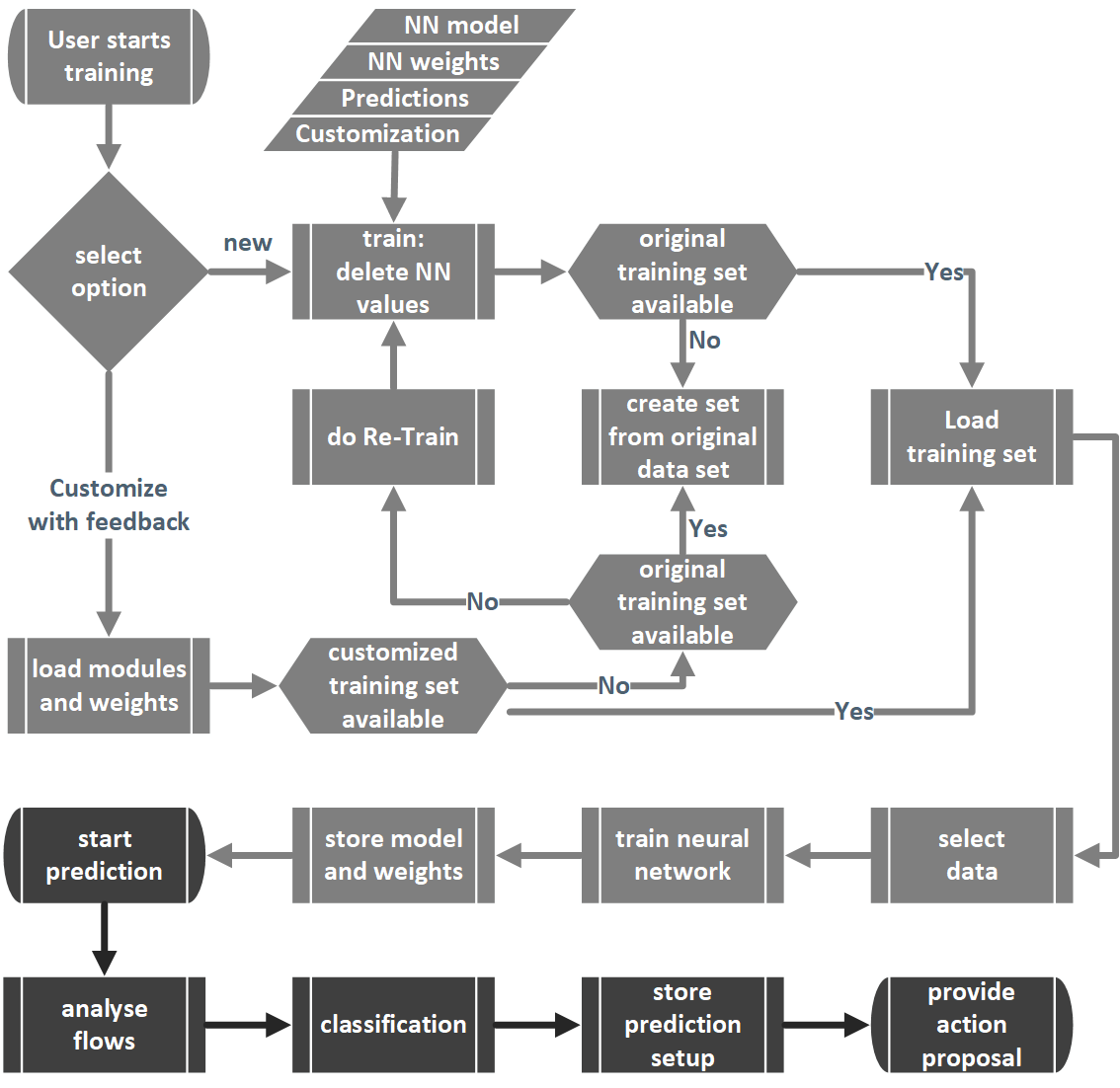}
	\caption{Training process and case prediction with the neural network for classification and customized suggestions based on labelled flow data.}
	\label{fig:trainingmode}
	\vspace*{-0.2cm}
\end{figure}

The process of the training is shown in Figure~\ref{fig:trainingmode}.
After a successful check, the set is loaded and the training is performed with a subset of the data.
After that the changed weights of the net will be saved in a suitable form.
If the check reveals that an unexpected deletion of the set has taken place, the entire original set is used to carry out the training in Retrain mode.

\Section{Evaluation}
To evaluate the concept, a first prototype, which is based on a multilayer perceptron emulated by \textit{Tensorflow} and \textit{Keras}, is implemented.
Different use cases were tested in regard to the user feedback and the feedback usage inside the training module. 

\begin{figure}[htb]
	\centering
	\includegraphics[width=0.48\textwidth]{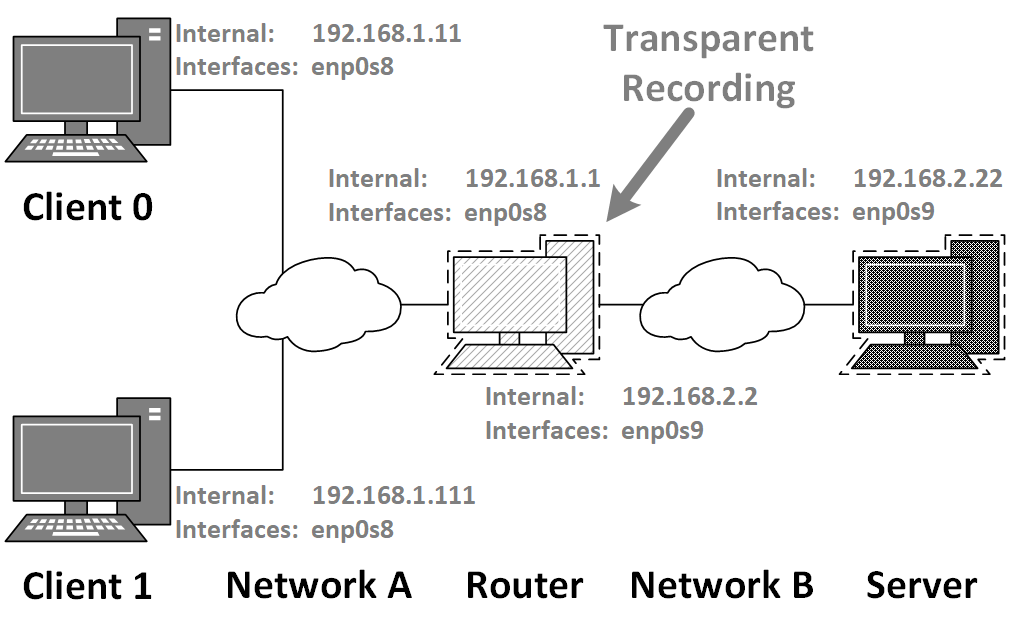}\\
	\caption{Structure of the test environment}
	\label{fig:Testumgebung}
	\vspace*{-0.2cm}
\end{figure}

\SubSection{Test environment}
The composition of the test laboratory, which is used to create the datasets for the different use cases used by the AI, is shown in Figure~\ref{fig:Testumgebung}. The server is providing the user with web services like \textit{NGINX} or \textit{Apache~2}. Simultaneously, the test environment is used to simulate user access, DoS attacks and Curl/Ping requests on the web service.
All DoS attacks were made with help of the \textit{Xerxes} and \textit{Torshammer} tools.
The so created network traffic is then transferred to the NERD system. 

\SubSection{Results and Valuation}
In order to analyse the network traffic there are three prioritised attributes out of the data template, which are selected and used to deliver the needed data sets to the AI.

First of all, there is the \textit{TCP-Rate} which helps to determine the varying behaviour of the TCP connection requests. Secondly, the \textit{TCPPAckCntAsm} shows asymmetries inside the data sets, which are then used to differentiate between regular data traffic and, for example, traffic created by DoS attacks. Lastly, there is \textit{perPS}, which shows the relationship between sent packages in a flow, allowing an efficient classification of the data in respect to different the behaviour patterns of the tested use cases. 

Based on these parameters, the AI calculates a probability for the possible outcome classes. Table~\ref{tabohnefeedback} shows the calculated values for six scenarios. 

\begin{table}[htb]
	\centering
	\caption{Results of prediction process in percent}
	\label{tabohnefeedback}
	\begin{tabular}{c|c|r|r|r}
		Data  & Class  & Traffic & Incident & Thread   \\ \hline
		1         & Web    & 0,76        & 99,23            & 0,00  \\
		2         & Web    & 32,57       & 67,19            & 0,24  \\
		3         & DoS    & 0,00        & 0,00             & 100,0 \\
		4         & DoS    & 0,00        & 0,00             & 100,0 \\
		5         & Normal & 100,00      & 0,00             & 0,0   \\
		6         & Normal & 99,98       & 0,02             & 0,0   \\
	\end{tabular}
	\vspace*{-0.2cm}
\end{table}

These results clearly show, that the classes, which were determined by the AI, meet the expected values, which therefore proves that pattern recognition for the used data was successfully implemented. 

\SubSection{User Feedback Usage}
At the beginning of the test, the NN is set to the same state of training as it was in the previous pattern recognition test. Based on the user feedback, which gets generated with the help of the rating module, the NN can learn from previous situations and events, which leads to more accurate and precise solutions. For this test, the user rated the class correlating to the expected outcome with the highest possible user rating. Table~\ref{tabfeedback} shows the resulting outcome of the following prediction. 

\begin{table}[htb]
	\centering
	\caption{Results of the prediction process after the user rating in percent}
	\label{tabfeedback}
	\begin{tabular}{c|c|r|r|r}
		Data  & Class  & Traffic & Incident & Thread   \\ \hline
		1         & Web    & 0,07        & 99,92            & 0,01  \\
		2         & Web    & 0,47        & 99,49            & 0,04  \\
		3         & DoS    & 0,00        & 0,00             & 100,0 \\
		4         & DoS    & 0,00        & 0,00             & 100,0 \\
		5         & Normal & 100,00      & 0,00             & 0,0   \\
		6         & Normal & 100,00      & 0,00             & 0,0   \\
	\end{tabular}
	\vspace*{-0.2cm}
\end{table}
It clearly shows a climb of approximately 32 \% for the Web 2 case in relation to the previous prediction. This shows a clear influence of the user input on the result set and a way to improve further predictions of the AI. 

\SubSection{Decision and Management Support}
The requirement for decision support is met by the PDF-Print module.
For every prediction made by the AI, a PDF file is created which contains all results calculated by the AI.
Additionally, it contains a summary of possible causes with their associated probabilities, as well as action proposals regarding hardware and software level. It can be taken as first step for documentation in a forensic process.
There is a separate specification for every single cause, so that further non-technical information can be included.
Solution measures contain different approaches depending on the cause, for example, dedicated information to firewall rulesets, a reboot of the affected servers when there is a service breakdown, or temporal blacklisting of specific IP addresses while notifying the service provider in case of a DoS attack.

Within the tests of the six aforementioned scenarios, the system always put out the appropriate and expected troubleshooting solution with the highest probability.
In addition, information is gathered on the organisation-specific knowledge base and pages are linked in the own wiki and confluence system.
Furthermore, a correlation to current attack vectors and vulnerabilities is made.
The automatically created reports also include best practices for the identified situation to be used by the higher management levels.

\Section{Conclusions}

In this paper, we presented the novel approach of NERD, a Cyber Incident Handling Support System.
It uses the information and technology of current state of the art Security Information and Event Management systems.
Therefore, attributes like byte asymmetry and syn-package ratios are used to identify and to classify potential incidents.
Combined with artificial intelligence, NERD goes one step further in the direction of self-healing systems and infrastructures.
It enables the analysation of data to provide decision and management support for Board of Directors via solutions and reports.
It suggests the technical user with concrete and detailed activities to handle efficiently an incident to reach swifter the normal status of operation than without our system.
Therefore, it uses a common knowledge base and best practises as well as information of past incidents and the specific setup of the company to provide multiple possibilities.
The artificial intelligence enables the system to be customized for future decision-making via feedback to be more effective by using labelled flow data as input for the learning process.. 
All tests with our prototype show an adequate and appropriate reaction of the system.
The usage of user feedback has been realised by implementing a user rating system for the suggested solutions.
All decisions made by the user and their effectiveness were then included in training the AI, which led to an improvement of the system.

The developed approach has a modular structure, shows a high degree of automation and can be easily expanded.
An improvement to recognition accuracy and interference elimination will be part of future work.
We will include more categories of incidents as well as more solution possibilities to build up a larger knowledge base.
Furthermore, we will include a functionality so share incident handling information in an anonymized way based in a common data exchange format~\cite{Steinberger2015}.
This allows the distribution of the incident handling knowledge and a quicker recovery.

\bibliographystyle{ieeetr}
\bibliography{ms,lniguide}

\end{document}